\documentstyle[11pt]{article}

\def\bra#1{{\langle#1|}}
\def\ket#1{{|#1\rangle}}

\def\e{{\rm e}}

\def\id{{\hat 1}}
\def\tr{{\rm Tr}}

\def\ent{{\cal E}}
\def\ebar{{\bar{\cal E}}}
\def\entp{{\cal E}^p}
\def\entf{{\cal E}^f}

\begin{document}

\title{Remotely prepared entanglement:  a quantum web page}

\author{Todd A. Brun\thanks{Email:  tbrun@ias.edu.  Phone:  (609) 734-8335.
FAX:  (609) 951-4489.} \\
Institute for Advanced Study, Einstein Drive, Princeton, NJ  08540}

\date{5 July 2001}

\maketitle

\begin{abstract}
In quantum teleportation, an unknown quantum state is transmitted from
one party to another using only local operations and classical
communication, at the cost of shared entanglement.  Is it possible similarly,
using an $N$ party entangled state, to have the state retrievable by
{\it any\/} of the $N-1$ possible receivers?  If the receivers cooperate,
and share a suitable state, this can be done reliably.  The $N$ party
GHZ is one such state; I derive a large class of such states, and show that
they are in general not equivalent to the GHZ.  I also briefly discuss
the problem where the parties do not cooperate, and the relationship
to multipartite entanglement quantification.  I define a new set of
entanglement monotones, the entanglements of preparation.
\end{abstract}


\section{Introduction}

Much of the recent progress in quantum information theory has stemmed
from the idea of using quantum entanglement as a resource for carrying
out operations on quantum states.  In particular, shared entanglement
between two parties can be used to {\it teleport\/} an unknown quantum
state using only local operations (measurements and unitary transformations)
and classical communication (LOCC) \cite{Bennett93}; it can also enable
two classical bits to be transmitted by a single quantum bit ({\it q-bit\/})
via {\it superdense coding\/} \cite{Bennett92}.

The accepted unit of quantum entanglement (at least for pure states) is
the maximally-entangled (EPR) pair, or {\it e-bit\/}.  Any such pair is
equivalent under local unitary transformations to the state
\begin{equation}
\ket{\Psi_{AB}} = (\ket{00} + \ket{11})/\sqrt{2} \;.
\end{equation}
The entanglement of any bipartite pure state can be measured in e-bits,
and in the asymptotic limit (i.e., where the two parties have many copies
of a state) any two states with the same entanglement can be reversibly
interconverted \cite{Bennett96a}.

By contrast, the case with more than two parties is far from clear.  No
single number suffices to describe the entanglement of multipartite
states, even asymptotically; nor is it known how many numbers are needed,
or even if the number is finite; nor are general algorithms known for
reversibly interconverting states \cite{Bennett99}.
In the face of these limits, it seems
reasonable both to try to generalize bipartite results to multipartite
systems, and to relate multipartite entanglement to the bipartite measure.

\section{The quantum web page}

Let us briefly review the teleportation protocol between two parties,
usually known as Alice and Bob.  It is assumed that they share a
maximally entangled pair of q-bits (one e-bit).  In addition, Alice has
another q-bit in an unknown state $\ket\chi$.  She jointly measures this
q-bit together with her half of the entangled pair in the Bell basis,
\begin{equation}
\ket{\phi^{\pm}} = (\ket{00} \pm \ket{11})/\sqrt{2} \;,\ \ 
\ket{\psi^{\pm}} = (\ket{01} \pm \ket{10})/\sqrt{2} \;,
\end{equation}
and transmits the result of her measurement (2 classical bits) to Bob.
This measurement of course destroys Alice's copy of the state she wishes
to send.  By performing one of four unitary transformations on his half
of the entangled pair, Bob can reconstruct the state $\ket\chi$
\cite{Bennett93}.

The problem I consider is a simple generalization of teleportation to $N$
parties.  Suppose that these $N$ parties (Alice, Bob, Cara,
David\ldots, Nancy) share an $N$ q-bit pure state.  One of them, say Alice,
is given another q-bit in an unknown state.  She wishes to combine this
state with the shared state in such a way that any of the other $N-1$
parties can retrieve it using only LOCC.  If we make an analogy between
ordinary teleportation and an email, this procedure would be more like a
{\it quantum web page}:  making a state available to anyone
in a given network.

Suppose that one party, say Bob, decides to retrieve the state.  He clearly
will only be able to do so with the cooperation of the other $N-2$ parties.
If these parties do not assist, he will at best share an entangled mixed
state with Alice, which is insufficient for reliable teleportation to
be carried out.  An easy way to see this is to note that if Bob could
retrieve the state without the assistance of the others, so could Cara
or any other party; and the state would have been copied, which is forbidden
by the no-cloning theorem \cite{Wootters82}.

If the other parties cooperate the procedure can indeed be carried out.
Suppose that they share an $N$ party GHZ state,
\begin{equation}
\ket{\Psi_{A\cdots N}} = ( \ket{0\cdots0} + \ket{1\cdots1} )/\sqrt{2}.
\end{equation}
These states are often considered to be {\it maximally entangled}, though
in the absence of a measure of multipartite entanglement it is not clear
that this makes sense \cite{Higuchi00}.
Let the parties C\ldots N measure their q-bits in the basis
\begin{equation}
\ket{0'} = (\ket0 - \ket1)/\sqrt{2},\ \ 
  \ket{1'} = (\ket0 + \ket1)/\sqrt{2} \;.
\end{equation}
After the first party measures his or her q-bit, the remaining $N-1$ parties
will share an $N-1$ party GHZ state.  This becomes an $N-2$ party GHZ with the
next measurement, and so on, until Alice and Bob are left with the state
\begin{equation}
\ket{\Psi_{AB}} = (\ket{00} + (-1)^{p(s)} \ket{11})/\sqrt{2} \;,
\end{equation}
where $s$ denotes the sequence of $N-2$ measurement results, and $p(s)$
is the parity of the sequence $s$.  This is a maximally entangled pair,
and can be used to teleport the unknown state from Alice to Bob.

Since the order in which these operations are performed doesn't matter,
Alice can carry out her part of the teleportation protocol before any
of the other measurements are carried out, even before it has been
decided which of the $N-1$ parties will retrieve the state.  She jointly
measures in the Bell basis the q-bit whose state she wishes to make available,
together with her part of the $N$ bit state, and broadcasts the measurement
result to the other parties.  At that point, any of them can retrieve
the state with the cooperation of the others, with no further action on
the part of Alice; indeed, Alice no longer shares entanglement with the
other parties.

We have seen how this can be done using an $N$ party GHZ state, but is
this the only state which will work?  Any state which allows a
maximally-entangled pair to be prepared between Alice and any other party,
using only LOCC, could be used instead.  I will call all such states
{\it web states\/}, and call the basis in which each party
measures his or her bit the {\it preparation basis\/}.  For now, I
restrict myself to the case where the choice of this basis does not
depend on which two parties are to share the pair, nor on the outcome
of measurements by other parties.  I will call this case {\it context-free\/}.

If we rewrite the GHZ in terms of the preparation bases of all the parties,
it takes the form
\begin{equation}
\ket{\Psi_{A\cdots N}} = {1\over\sqrt{2^{N-1}}} \sum_{{\rm even}\ s}
 \ket{s} \;,
\end{equation}
where $s$ is a sequence of $N$ 0's and 1's, and the sum is restricted only
to those with an even number of 1's.  In this form it is easy to see that
any measurement of $N-2$ of the bits in their preparation bases will leave
the other two in a maximally-entangled state.

Are all such context-free web states equivalent to the $N$ party GHZ?  No,
they are not.  Consider the following state:
\begin{equation}
\ket{\Psi_{A\cdots N}} = {1\over\sqrt{2^{N-1}}} \sum_{{\rm even}\ s}
 \exp(i\theta(s)) \ket{s} \;,
\label{web_state}
\end{equation}
where the $\theta(s)$ are $2^{N-1}$ arbitrary phases.  Just as with the
GHZ, measuring any $N-2$ of the bits in this basis will leave the other
two in a maximally entangled state, so this state can be used for a
quantum web page.  But a state of this form cannot, in general, be
transformed into a GHZ by LOCC.  There are $2^{N-1}$ phases; at most $N+1$
of these can be eliminated by local unitary transformations.  Thus, such
states can only be transformed into the GHZ in general for $N=3$.

In the next section, I will prove that all web states with context-free
preparation bases are equivalent to a state of the form (\ref{web_state})
under local unitary transformations.  I will also prove that if a state
allows a maximally entangled pair to be prepared between one particular
party (say Alice) and any of the others, then it allows a maximally entangled
pair to be prepared between {\it any\/} two parties, and hence can be put
in the form (\ref{web_state}).

\section{$N$ q-bit web states}

Suppose we select two parties (Alice and Bob) for whom we wish to prepare
an EPR pair.  The other $N-2$ parties measure their bits in the correct
preparation basis, producing a measurement results $s$ (where $s$ denotes
a sequence of $N-2$ 0's and 1's).  For any outcome $s$, Alice and Bob must
be left with a state $\ket{\Psi_{AB}(s)}$ which is maximally entangled.
So the full state can be written
\begin{equation}
\ket{\Psi_{A\cdots N}} = \sum_s c_s \ket{\Psi_{AB}(s)} \otimes \ket{s} \;,
\label{state_form1}
\end{equation}
where the $c_s$ are real, positive coefficients with normalization
\begin{equation}
\sum_s c_s^2 = 1 \;.
\end{equation}
The most general form of a maximally entangled pair is
\begin{eqnarray}
\ket{\Psi_{AB}} &=& \cos\alpha \left( \e^{i\theta} \ket{00} +
  \e^{i\phi} \ket{11} \right)/\sqrt{2} \nonumber\\
&&  + \sin\alpha \left( \e^{i\omega} \ket{01} -
  \e^{i(\theta+\phi-\omega)} \ket{10} \right)/\sqrt{2} \;,
\label{maximally_entangled}
\end{eqnarray}
where $0 \le \alpha \le \pi/2$ and $\theta,\phi,\omega$ are arbitrary phases.
One can derive this form straightforwardly, by writing down an arbitrary
two-bit pure state and imposing the restriction that
$\tr_B\{\ket{\Psi_{AB}}\bra{\Psi_{AB}}\} = \id/2$; this lets one eliminate
three degrees of freedom, yielding (\ref{maximally_entangled}).
We assume without loss of generality that q-bit B is also written in its
preparation basis.
We can now write the state (\ref{state_form1}) in the form
\begin{eqnarray}
\ket{\Psi_{A\cdots N}} &=& {1\over\sqrt2} \sum_s c_s \biggl[
  \cos\alpha_s \left( \e^{i\theta_s} \ket{00} +
  \e^{i\phi_s} \ket{11} \right) \nonumber\\
&& + \sin\alpha_s \left( \e^{i\omega_s} \ket{01} -
  \e^{i(\theta_s+\phi_s-\omega_s)} \ket{10} \right)
  \biggr] \otimes \ket{s} \;.
\label{state_form2}
\end{eqnarray}

For this to be a web state, we must be able to exchange Bob's q-bit
with that of any other party and recover a state of form
(\ref{state_form2}).  Suppose we wish to exchange Bob's bit with that
of the $M$th party.  Let us group all sequences $s$ into pairs
$(s_0,s_1)$ which are identical except at the $M$th bit, which is 0 for $s_0$
and 1 for $s_1$.  If we exchange Bob's q-bit with the $M$th q-bit,
the state becomes
\begin{eqnarray}
\ket{\Psi_{A\cdots N}} &=& {1\over\sqrt2} \sum_{(s_0,s_1)} \biggl[ \bigl(
  c_{s_0} \cos\alpha_{s_0} \e^{i\theta_{s_0}} \ket{00} -
  c_{s_0} \sin\alpha_{s_0}
  \e^{i(\theta_{s_0}+\phi_{s_0}-\omega_{s_0})} \ket{10} \nonumber\\
&& + c_{s_1} \cos\alpha_{s_1} \e^{i\theta_{s_1}} \ket{01} -
  c_{s_1} \sin\alpha_{s_1}
  \e^{i(\theta_{s_1}+\phi_{s_1}-\omega_{s_1})} \ket{11} \bigr)
  \otimes \ket{s_0} \nonumber\\
&& + \bigl( c_{s_0} \cos\alpha_{s_0} \e^{i\phi_{s_0}} \ket{10} +
  c_{s_0} \sin\alpha_{s_0} \e^{i\omega_{s_0}} \ket{00} \nonumber\\
&& + c_{s_1} \cos\alpha_{s_1} \e^{i\phi_{s_1}} \ket{11} +
  c_{s_1} \sin\alpha_{s_1} \e^{i\omega_{s_1}} \ket{01} \bigr)
  \otimes \ket{s_1} \biggr] \;.
\label{exchange}
\end{eqnarray}
For this to be a web state, there must be new parameters
$c'_s,\alpha'_s,\theta'_s,\phi'_s,\omega'_s$ which put (\ref{exchange})
in form (\ref{state_form2}).

Comparing (\ref{exchange}) and (\ref{state_form2}) term by term, we
immediately see that we must have
\begin{eqnarray}
c'_{s_0}\cos\alpha'_{s_0} &=& c_{s_0}\cos\alpha_{s_0}
  = c_{s_1}\sin\alpha_{s_1} \;, \nonumber\\
c'_{s_0}\sin\alpha'_{s_0} &=& c_{s_0}\sin\alpha_{s_0}
  = c_{s_1}\cos\alpha_{s_1} \;, \nonumber\\
c'_{s_1}\cos\alpha'_{s_1} &=& c_{s_0}\sin\alpha_{s_0}
  = c_{s_1}\cos\alpha_{s_1} \;, \nonumber\\
c'_{s_1}\sin\alpha'_{s_1} &=& c_{s_0}\cos\alpha_{s_0}
  = c_{s_1}\sin\alpha_{s_1} \;.
\end{eqnarray}
This must hold for every pair $(s_0,s_1)$, and for every choice of $M$,
which implies that
\begin{equation}
c_s = 1/\sqrt{2^{N-2}} \ \ \forall s \;,
\end{equation}
and that
\begin{equation}
\alpha_s = \alpha_{p(s)} \ \ \forall s \;,\ \ 
  \alpha_1 = \pi/2 - \alpha_0 \;,
\end{equation}
where $p(s)$ is the parity of the string $s$, with $p(s)=0$ for even strings,
$p(s)=1$ for odd strings.  This just means that
$\cos\alpha_0 = \sin\alpha_1$ and $\sin\alpha_0 = \cos\alpha_1$.

In a similar way we can examine the phases of (\ref{exchange}). It is only
possible to preserve the form of (\ref{state_form2}) if
\begin{equation}
\exp i( \phi_{s_1} - \omega_{s_1} ) =
  - \exp i( \phi_{s_0} - \omega_{s_0} ) \;.
\end{equation}
Again, this must hold for any pair $(s_0,s_1)$, and for any choice of $M$,
which gives us the requirement
\begin{equation}
\omega_s = \phi_s + \gamma + p(s)\pi \;,
\end{equation}
where $\gamma$ is a constant phase, and $p(s)$ is once again the parity of
the string $s$.  We can eliminate the phase $\gamma$ by a local unitary
transformation
\begin{eqnarray}
\ket0_A \rightarrow \ket0_A \;, && \ket0_B \rightarrow \ket0_B \;, \nonumber\\
\ket1_A \rightarrow \e^{i\gamma} \ket1_A \;,
&& \ket1_B \rightarrow \e^{-i\gamma} \ket1_B \;,
\end{eqnarray}
which leaves the physical preparation basis of $B$ unchanged.

With these restrictions, we now see that any context-free web state can be
written in the form
\begin{eqnarray}
\ket{\Psi_{A\cdots N}} &=& {1\over\sqrt{2^{N-1}}} \sum_s \biggl[
  \cos\alpha_{p(s)} \left( \e^{i\theta_s} \ket{00} +
  \e^{i\phi_s} \ket{11} \right) \nonumber\\
&& + (-1)^{p(s)} \sin\alpha_{p(s)} \left( \e^{i\phi_s} \ket{01} -
  \e^{i\theta_s} \ket{10} \right)
  \biggr] \otimes \ket{s} \;.
\label{state_form3}
\end{eqnarray}
But is this the simplest form possible?  Let us perform a local unitary
transformation on q-bit A:
\begin{eqnarray}
\ket0_A &\rightarrow&
  \cos\alpha_0 \ket0_A + \sin\alpha_0 \ket1_A \;, \nonumber\\
\ket1_A &\rightarrow&
  - \sin\alpha_0 \ket0_A + \cos\alpha_0 \ket1_A \;.
\end{eqnarray}
This must still be a web state, since a local unitary on A will leave
any maximally entangled pair still maximally entangled.  We can easily
check that the state (\ref{state_form3}) now takes the form
\begin{equation}
\ket{\Psi_{A\cdots N}} = {1\over\sqrt{2^{N-1}}} \sum_s \left\{
  { {\exp(i\theta(s)) \ket{00} + \exp(i\phi(s)) \ket{11}} \atop
  {\exp(i\theta(s)) \ket{10} - \exp(i\phi(s)) \ket{01}}}
  \right\} \otimes\ket{s} \;,
\label{state_form4}
\end{equation}
where the upper (lower) line is used for even (odd) strings $s$.
This expression is exactly equivalent to the form (\ref{web_state})
given in section 2.  So we see that {\it any\/} context-free web state
can be put in form (\ref{web_state}), and any state which allows context-free
preparation of an EPR between one party and any of the others must
allow preparation of an EPR between {\it any\/} two parties.

\section{Generalizations}

\subsection{Contextual web states}

The most obvious generalization we might consider is to states which
are not context-free, i.e., where the choice of measurement basis
depends on which two parties are to have an EPR pair prepared.  Logically
this should be a much larger class of states than that of context-free
web states.

I have not derived a general prescription for all contextual web states,
but some obvious general characteristics suggest themselves.  Any web state
must give a reduced density matrix proportional to the identity for all
q-bits:
\begin{equation}
\tr_{N-1} \left\{ \ket{\Psi_{A\cdots N}}\bra{\Psi_{A\cdots N}} \right\}
  = \rho = \id/2 \;.
\end{equation}
Also, any {\it pair\/} of q-bits A and M must have a reduced density matrix
$\rho_{AM}$ with entanglement of assistance (or hidden entanglement) equal
to 1 \cite{Cohen98,DiVincenzo98}.  These are necessary
but probably not sufficient conditions.  For $N=3$, these requirements
are enough to prove that {\it all\/} web states are equivalent to the GHZ
under local unitary transformations, and hence are context-free.

For $N=4$ we can find examples which are not equivalent to context-free
states, and hence cannot be written in the form (\ref{web_state}).  For
instance, the state
\begin{equation}
\ket{\Psi_{ABCD}} = {1\over2} \biggl[
  \left(\ket{00} + \ket{11} \right) \otimes \ket{00}
  + \left(\ket{00} - \ket{11} \right) \otimes \ket{11} \biggr] \;,
\label{contextual}
\end{equation}
allows an EPR to be prepared between A and B or between C and D
by measuring the other two bits in the given basis.  However, by measuring
in that basis it is impossible to prepare an EPR between any other pair.

We could instead measure two bits in the basis
\begin{eqnarray}
\ket{0'} &=& \left( \ket0 + \ket1 \right)/\sqrt2 \;, \nonumber\\ 
\ket{1'} &=& \left( \ket0 - \ket1 \right)/\sqrt2 \;.
\label{xbasis}
\end{eqnarray}
A maximally-entangled state can be prepared between A and C, B and C,
A and D or B and D by measuring the other two bits in the basis
(\ref{xbasis}), but {\it not\/} between A and B or C and D.
This state (\ref{contextual}) is thus a web state, but a contextual one.

An even more general possibility would be to allow the measurement bases
of some bits to depend on measurement outcomes for others; or to allow
general positive-operator valued measurements (POVMs); or both.
Characterizing this large class of operations, though, is not easy
\cite{Bennett98}; no simple form is known for a general LOCC procedure.

\subsection{Preparing other states}

We might also ask about preparing shared states other than EPR pairs.
Because an EPR pair can be reliably transformed into any two q-bit state
by LOCC \cite{Nielsen99},
the web states we have considered can also produce any such state between
any two parties.

For multipartite states the situation is less clear, but some results follow
easily.  Note that if one party carries out a measurement on an $N$ party
context-free web state in his or her preparation basis,
the other $N-1$ parties are left with an $N-1$ party context-free web state.
For $N=3$, all such states are equivalent to the 3 party GHZ.  So for
$N>3$, any state of form (\ref{web_state}) also allows the preparation of
a GHZ among any three parties.

For more general cases, a procedure like that of section 3 might be
carried out to find all $N$ party states which allow some particular $M<N$
party state to be reliably prepared among any $M$ participants.  But
there seems no guarantee that the results will in general be as simple
as those for the EPR pair.

\subsection{The noncooperative problem and cloning}

This quantum web page algorithm requires that all the parties cooperate
in enabling the selected party (say Bob) to retrieve the state
$\ket\chi$ that Alice has made available.
Without this cooperation, at best Bob will retrieve a mixed state.

However, we can imagine a version of this problem in which the parties
do {\it not\/} cooperate, and ask what state should be used to maximize
the fidelity of Bob's state $\rho_B$ with the state $\ket\chi$ he is
attempting to retrieve.  Let $\ket{\Psi_{A\cdots N}}$ be the shared initial
state, and $\ket\chi$ be the state that Alice is trying to make available.
After Alice carries out her part of the procedure (which is just like her
part of the cooperative procedure), the other $N-1$ parties are left with a
new joint state $\ket{\Psi'_{B\cdots N}}$.  The reduced state of Bob's
subsystem is
\begin{equation}
\rho_B = \tr_{C\cdots N} \left\{ \ket{\Psi'_{B\cdots N}}\bra{\Psi'_{B\cdots N}}
  \right\} \;,
\end{equation}
and we wish to maximize the fidelity
\begin{equation}
F = \bra\chi \rho_B \ket\chi \;.
\label{cloning_fidelity}
\end{equation}

This fidelity should be the same for any party and for any state $\ket\chi$,
so we see that this is the same as $N-1$ party {\it symmetric cloning\/}
\cite{Cloning}.  The maximum fidelity achievable by symmetric cloning
to $N-1$ q-bits is $F=(2N-1)/(3(N-1))$.  The class of states which are
used in the noncooperative case is quite different from the class of
web states; examples which achieve this maximum fidelity have been
discovered, e.g., by D\"ur \cite{Dur01} and Koashi et al. \cite{Koashi00}.
A related problem has also been considered by D\"ur and Cirac \cite{Dur00}.

Note that one must not confuse the {\it cloning fidelity} $F$ given by
(\ref{cloning_fidelity}) with the {\it singlet fraction} $F_s$ of the
$N$-partite state.  If we trace out parties C\ldots N of
$\ket{\Psi_{A\cdots N}}$
to get the bipartite density matrix $\rho_{AB}$, the singlet fraction is
\begin{equation}
F_s = \max_{\ket{\Psi_{AB}}} \bra{\Psi_{AB}}\rho_{AB}\ket{\Psi_{AB}} \;,
\label{singlet_fraction}
\end{equation}
where the maximum is over all {\it maximally entangled states}
$\ket{\Psi_{AB}}$.  The maximum of $F_s$ is considerably lower than $F$,
and is bounded by $F_s \le N/2(N-1)$.  This distinction is brought out
in \cite{Horodecki98}.

\subsection{Multipartite entanglement}

The problem of characterizing multipartite entanglement remains largely
unsolved, especially in the asymptotic limit \cite{Bennett99}.
The idea behind these web
states suggests a set of possibly useful quantities that can be calculated
from a multipartite pure state.

The {\it entropy of entanglement\/} for an entangled pair is
\begin{equation}
\ent(\ket{\Psi_{AB}}) = - \tr_A \{ \rho_A \log_2 \rho_A \} \;,
\end{equation}
where $\rho_A$ is the reduced density matrix of q-bit A,
\begin{equation}
\rho_A = \tr_B\{ \ket{\Psi_{AB}}\bra{\Psi_{AB}} \} \;.
\end{equation}
Suppose we perform some LOCC procedure (e.g., a measurement) on
an $N$-partite system which is initially in a pure state
$\ket{\Psi_{A\cdots N}}$; denote this procedure by $O$, with outcomes
$o_i$ occurring with probability $p_i$, and assume that it leaves the
pair of subsystems A and B in a {\it pure} state $\ket{\Psi_{AB}(o_i)}$.
(That is, there is no entanglement left with C\ldots N.)  Then we can
define the {\it average preparation entanglement\/} of A and B under $O$:
\begin{equation}
\ebar_{AB}(\ket{\Psi_{A\cdots N}},O) =
  \sum_i p_i \ent(\ket{\Psi_{AB}(o_i)}) \;.
\label{ebar_def}
\end{equation}

This quantity will vary enormously depending on the choice of $O$.  We can,
however, look for the maximum, and use that to define the {\it entanglement
of preparation\/}
\begin{equation}
\entp_{AB}(\ket{\Psi_{A\cdots N}}) \equiv
  \max_O \ebar(\ket{\Psi_{A\cdots N}},O) \;.
\label{entanglement_preparation}
\end{equation}
For an $N$-partite system, we can define the entanglement of preparation
between any two parties.  This quantity is similar to the entanglement
of assistance (or hidden entanglement) $\ent^a(\rho_{AB})$
\cite{Cohen98,DiVincenzo98}; however, in general it will be lower,
$\entp_{AB}(\ket{\Psi_{A\cdots N}}) \le \ent^a(\rho_{AB})$,
since the measurements yielding the entanglement of preparation must respect
the division of C\ldots N into local subsystems, while the entanglement of
assistance has no such constraint.  By definition, any web state must have
entanglement of preparation 1 between q-bit A and any other q-bit.

Note that we can relax the restriction on $O$ (to operations which leave
A and B in a pure state) by replacing the entropy of entanglement in
(\ref{ebar_def}) with the {\it entanglement of formation}
$\entf(\rho_{AB}(o_i))$ \cite{Bennett96b}.  With this change, the maximum in
(\ref{entanglement_preparation}) becomes a maximum over {\it all}
LOCC procedures $O$.
It is not hard to see that the value of $\entp_{AB}(\ket{\Psi_{A\cdots N}})$
is unchanged by this alteration; the maximum will always be achieved by
an operation $O$ which {\it does} leave A and B in a pure state, for which
the entropy of entanglement and the entanglement of formation give
equal values.

An important point is that, by construction,
$\entp_{AB}(\ket{\Psi_{A\cdots N}})$ is an {\it entanglement monotone}.
That is, the value of $\entp_{AB}$ cannot increase on average if one
performs some LOCC procedure on the state $\ket{\psi_{A\cdots N}}$.  Clearly,
if it {\it could} increase, then the operation $O$ would not be the maximum.
Such monotones are important in trying to understand the entanglement
properties of multipartite states.  In a similar way, any numerical function
of the state can be used to define an entanglement monotone; one simply
takes the average value of the function under the action of some LOCC
procedure $O$, and maximizes over all $O$.  Of course,
$\entp_{AB}(\ket{\Psi_{A\cdots N}})$ may be difficult to calculate exactly for
a general state; however, one may still be able to derive useful bounds on it.

Finally, the entanglement of preparation is a {\it superadditive} quantity.
Suppose that the parties A\ldots N share {\it two} systems with $N$
subsystems each, the two systems being in pure states $\ket{\Psi_{A\cdots N}}$
and $\ket{\Phi_{A\cdots N}}$, respectively.  We can consider these jointly
as a single $N$-partite system in state
$\ket{\Psi_{A\cdots N}}\otimes\ket{\Phi_{A\cdots N}}$.  Then the entanglement
of preparation for this joint state must satisfy
\begin{equation}
\entp_{AB}(\ket{\Psi_{A\cdots N}}\otimes\ket{\Phi_{A\cdots N}}) \ge
\entp_{AB}(\ket{\Psi_{A\cdots N}}) + \entp_{AB}(\ket{\Phi_{A\cdots N}}) \;.
\end{equation}
This is similar to the class of entanglement monotones derived in
\cite{Barnum01}.

\section{Conclusions}

It is possible to produce a quantum analogue of a web page:  a
procedure by which a quantum state is made available by one party to any
of a group of others, using only a shared initial state,
local operations and classical communication.  Unlike the classical case,
however, retrieving the quantum state requires cooperation among the
$N-1$ possible recipients, and only one can actually retrieve it.

I have presented a large class of $N$ q-bit pure states which can be
used for the quantum web page, of which the $N$ party GHZ is an example,
and argued that an even larger class is potentially available.
This problem suggests a set of quantities,
the {\it entanglements of preparation\/} of the various pairs, which may
prove useful in the ongoing attempt to characterize multipartite entanglement,
especially since these quantities are entanglement monotones.
This problem could be generalized in many ways; but for the present,
these related problems remain unsolved.

\section*{Acknowledgments}

I would like to thank Oliver Cohen and Bob Griffiths for their
feedback on this problem, and for looking over the manuscript; and
Michele Mosca and Alain Tapp for inviting my contribution to this special
issue of {\sl Algorithmica}.

After this work was completed I became aware
of a paper by Briegel and Raussendorf \cite{Briegel01}, which includes
a definition of {\it maximal connectedness\/} which is the same as
my web page property; they also give a particular set of states which have
that property.  In particular, their state
$\ket{\phi_4}$ is equivalent to the contextual
web state (\ref{contextual}) in this paper.

This work was supported by
DOE Grant No.~DE-FG02-90ER40542.

\end{document}